# Electron-phonon coupling constant and BCS ratios in (La,Nd)-H superhydride

Evgueni F. Talantsev[1,2]

[1]M.N. Mikheev Institute of Metal Physics, Ural Branch, Russian Academy of Sciences, 18, S. Kovalevskoy St., Ekaterinburg, 620108, Russia

[2]NANOTECH Centre, Ural Federal University, 19 Mira St., Ekaterinburg, 620002, Russia

**Abstract**

Stoichiometric near-room temperature superconductors (NRTS) (for instance, $H_3S$ and $LaH_{10}$) exhibit a high ground state upper critical field, $B_{c2}(0) > 100$ T, such that the magnetic phase diagram in these materials cannot be measured in non-destructive experiments. However, Semenok *et al.* (2022 *arXiv*2203.06500) proposed idea of exploring the full magnetic phase diagram in NRTS samples, in which superconducting order parameter is suppressed by magnetic element doping. If the element is uniformly distributed in the material, then the theory of electron-phonon mediated superconductivity predicts the suppression of the order parameter in three-dimensional *s*-wave superconductor. Semenok *et al.* (arXiv2203.06500) experimentally proved this idea by substituting lanthanum with the magnetic rare earth neodymium in the $(La_{1-x}Nd_x)H_{10-y}$. As a result, the transition temperature in $(La_{1-x}Nd_x)H_{10-y}$ (x = 0.09) was suppressed to $T_c$~120 K, and the upper critical field decreases to $B_{c2}(T=41$ K$)=55$ T. While the exact hydrogen content should be further established in the $(La_{1-x}Nd_x)H_{10-y}$ (x = 0.09) (because similar $T_c$ suppression was observed in hydrogen deficient $LaH_{10-y}$ samples reported by Drozdov *et al* (2019 *Nature* **569** 528)), a significant part of the full magnetic phase diagram for $(La_{1-x}Nd_x)H_{10-y}$ (x = 0.09) sample was measured. Here we analyzed reported by Semenok *et al* (*arXiv*2203.06500) magnetoresistance data for $(La_{1-x}Nd_x)H_{10-y}$ (x = 0.09) compressed at $P$=180 GPa and deduced: (a) Debye temperature, $T_\theta = 1156 \pm 6\ K$, (b) the electron-phonon coupling constant, $\lambda_{e-ph} = 1.65 \pm 0.01$; (c) the ground state superconducting energy gap, $\Delta(0) =$ $20.2 \pm 1.3\ meV$; (d) the gap-to-transition temperature ratio, $\frac{2\Delta(0)}{k_B T_C} = 4.0 \pm 0.2$; and (e) the relative jump in specific heat at transition temperature, $\frac{\Delta C}{C} = 1.68 \pm 0.15$. The deduced values indicate that $(La_{1-x}Nd_x)H_{10-y}$ (x = 0.09; $P$ = 180 GPa) is a moderately strongly coupled s-wave superconductor.

**Electron-phonon coupling constant and BCS ratios in (La,Nd)-H superhydride**

**I. Introduction**

The discovery of a superconducting state with transition temperature above 200 K in highly compressed H$_3$S by Drozdov *et al.* [1] with the consequent discovery of high-temperature superconductivity in superhydrides of thorium [2] and near room temperature superconductivity (NRTS) in superhydrides of lanthanum [2,3], yttrium [4-6] and lanthanum-yttrium [7] represent the most fascinating scientific explorations in the field of superconductivity since the discovery of high-temperature superconductivity in cuprates [8].

While in the majority of theoretical works (comprehensive review has been published recently [9]) the electron-phonon mechanism is considered to be the primary pairing mechanism in NRTS superhydrides, alternative approaches to the nature of charge carrier interaction and pairing in NRTS are also under development [10-12]. One of the most solid experimental facts that supports the electron-phonon pairing mechanism in superhydrides is the prominent isotope effect with respect to the transition temperature, $T_c$ [1,3,13]. However, the effect of hydrogen-deuterium exchange on other fundamental parameters of NRTS superhydrides, for instance, on the lower and upper critical fields, as well as on Bardeen-Cooper-Schrieffer (BCS) ratios (i.e., $\frac{2\Delta(0)}{k_B T_c}$ where $\Delta(0)$ is the ground state of the superconducting energy gap, $k_B$ is the Boltzmann constant), remains to be explored.

Another important question which needs to be answered is the superconducting gap symmetry in NRTS hydrides. From the author's best knowledge, there is general agreement [9] that the superconducting energy gap in superhydrides exhibits *s*-wave symmetry. However, the first experimental evidence which confirms *s*-wave gap symmetry was only recently reported by Semenok *et al* [14]. This research group proposed to verify the *s*-wave gap symmetry in NRTS by employing one of the conclusions of Abrikosov-Gor'kov [15],

Anderson [16], and Openov [17,18] theories of dirty superconductors. This conclusion is that uniformly distributed (on the atomic level) impurities exhibited magnetic moment should suppress the superconducting order parameter in *s*-wave superconductors, but this kind of impurities should not affect the superconducting order parameter in *d*-wave superconductors. However, non-magnetic impurities should cause the suppression of in *d*-wave superconductors, but this kind of doping should not affect the *s*-wave superconducting state.

Thus, to reaffirm/disprove *s*-wave symmetry gap in LaH$_{10}$, Semenok *et al* [14] performed gradual doping of lanthanum decahydride by magnetic rare earth element, neodymium. The hydrogen/rare earth elements stoichiometry was kept 1/10 in all samples. While the exact hydrogen content in synthesized samples of La$_{1-x}$Nd$_x$H$_{10-y}$ (x = 0.08, 0.09, 0.20, 0.25, and 0.50) [14] should be further established (because similar $T_c$ suppression was observed in hydrogen deficient LaH$_{10-y}$ samples reported by Drozdov *et al* [3]), a significant part of the full magnetic phase diagram for (La$_{1-x}$Nd$_x$)H$_{10-y}$ (x = 0.09) sample was measured [14]. More specifically, it should be noted that there is a need for further experimental studies to confirm that all Nd atoms replace the lanthanum in their sites in the crystal lattice, instead than Nd will form the secondary phases, and that the hydrogen content remains to be stoichiometric.

Semenok *et al* [14] reported on gradual $T_c$ suppression on the increase in the Nd concentration. One of the interesting consequences associated with this $T_c$ suppression is that the upper critical field, $B_{c2}(T)$, is also decreasing. This makes it possible to measure $B_{c2}(T)$ for La$_{1-x}$Nd$_x$H$_{10}$ within a much wide reduced temperature range, $\frac{T}{T_c}$, in comparison with the range available for undoped stoichiometric NRTS materials H$_3$S [19] and LaH$_{10}$ [20], which exhibits the ground state upper critical field well above the value which can be measurable in non-destructive experiments.

Due to the upper critical field, $B_{c2}(T)$, is one of two fundamental fields in type-II superconductors, measured $B_{c2}(T)$ datasets can be used to extract several fundamental parameters of the superconductor [21,22], for instance:

1. ground state energy gap, $\Delta(0)$;
2. relative jump in electronic specific heat at $T_c$, $\Delta C/C$;
3. ground state coherence length, $\xi(0)$;
4. gap-to-transition temperature ratio, $\frac{2\Delta(0)}{k_B T_c}$ (where $k_B$ is the Boltzmann constant).
5. Fermi temperature, $T_F$

There are two other fundamental parameters of the electron-phonon mediated superconductors:

6. Debye temperature, $T_\theta$;
7. the electron-phonon coupling constant, $\lambda_{e-ph}$.

Debye temperature can be deduced from the fit of experimentally measured temperature dependent resistance, $R(T)$, to the Bloch-Grüneisen (BG) equation [23,24]:

$$R(T) = R_0 + A \left(\frac{T}{T_\theta}\right)^5 \int_0^{\frac{T_\theta}{T}} \frac{x^5}{(e^x - 1)(1 - e^{-x})} dx \qquad (1)$$

where $R_0$ and $A$ are free fitting parameters. From the deduced $T_\theta$ and measured $T_c$, the electron-phonon coupling constant, $\lambda_{e-ph}$, can be calculated as unique root of advanced McMillan equation [25,26]:

$$T_c = \left(\frac{1}{1.45}\right) \times T_\theta \times e^{-\left(\frac{1.04(1+\lambda_{e-ph})}{\lambda_{e-ph} - \mu^*(1+0.62\lambda_{e-ph})}\right)} \times f_1 \times f_2^* \qquad (2)$$

where

$$f_1 = \left(1 + \left(\frac{\lambda_{e-ph}}{2.46(1+3.8\mu^*)}\right)^{3/2}\right)^{1/3} \qquad (3)$$

$$f_2^* = 1 + (0.0241 - 0.0735 \times \mu^*) \times \lambda_{e-ph}^2. \qquad (4)$$

where $\mu^*$ is the Coulomb pseudopotential parameter, which can be assumed to be $\mu^* = 0.13$ for all NRST materials.

In this work, we deduced $\Delta(0)$, $\Delta C/C$, $\xi(0)$, $\frac{2\Delta(0)}{k_B T_c}$, $T_F$, $T_\theta$, and $\lambda_{e-ph}$ for La$_{1-x}$Nd$_x$H$_{10-y}$ (x = 0.09) compressed at pressure $P = 180\ GPa$ by analysing experimental $R(T)$ and $B_{c2}(T)$ datasets reported by Semenok et al [14,27]. Deduced parameters showed that La$_{1-x}$Nd$_x$H$_{10-y}$ (x = 0.09; $P = 180\ GPa$) is moderately strong coupled superconductor.

The fit of $R(T)$ dataset to Eq. 1 is shown in Fig. 1. This $R(T)$ dataset reported in Figure 2,a in the Ref. 14 and raw data file is freely available online by Semenok et al [27]. Deduced parameters are: $R_0 = 0.729(5)$, $A = 5.42 \pm 0.06$, and $T_\theta = 1156 \pm 6\ K$. By utilizing general requirement [26], that $T_c$ should be defined at the lowest as possible $\frac{R(T)}{R_{norm}(T)}$ ratio, and considering that the same criterion should be used to define $B_{c2}(T)$ dataset from $R(T, B)$ curves (reported in Figs. 2(b,c) [14]), the ratio of $\frac{R(T)}{R_{norm}(T)} = 0.08$ was used. In the result, $T_c$ for $R(T)$ in Fig. 1 was defined as $T_{c,0.08} = 122\ K$.

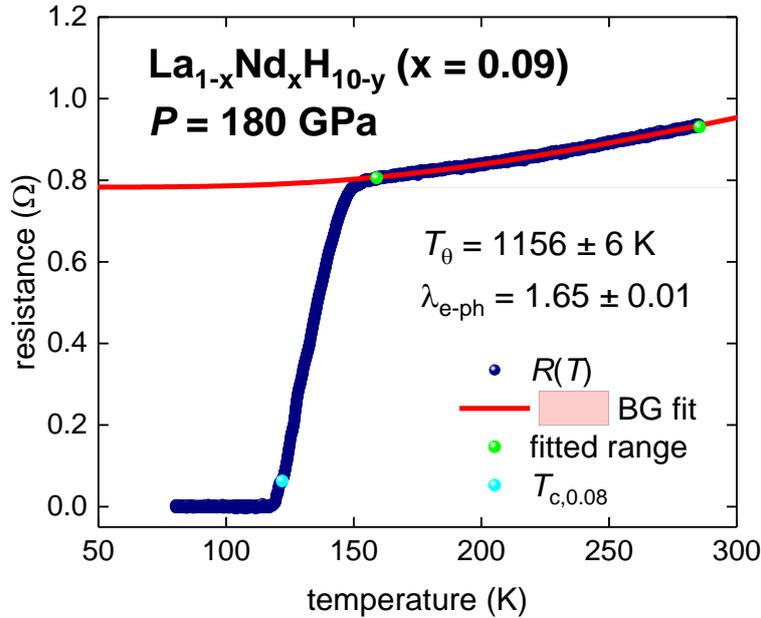

**Figure 1.** $R(T)$ data for highly compressed La$_{1-x}$Nd$_x$H$_{10-y}$ (x = 0.09; $P$ = 180 GPa) and data fit to Eq. 1 (raw data is freely available online by Semenok et al [14,27]). Green balls indicate the bounds for which $R(T)$ data was used for the fit to Eq. 1. Deduced $T_\theta = 1156 \pm 6\ K$, $T_{c,0.08} = 122\ K$, $\lambda_{e-ph} = 1.65 \pm 0.01$, fit quality is 0.9977. 95% confidence bands are shown by pink shadow areas.

The root of Eqs. 2-4 for given $T_\theta$, $T_c$, and $\mu^* = 0.13$ is $\lambda_{e-ph} = 1.65 \pm 0.01$. This deduced value is in a good agreement with $\lambda_{e-ph}$ values calculated by first-principles calculations by Semenok *et al* [27] in their Table S5.

By utilizing the same criterion of $\frac{R(T)}{R_{norm}(T)} = 0.08$, the $B_{c2}(T)$ dataset was derived from $R(T,B)$ curves showed in Figs. 2(b,c) of Ref. 14. In Fig. 2 the $B_{c2}(T)$ dataset is fitted to the equation for temperature dependent upper critical field for *s*-wave superconductors [21,22]:

$$B_{c2}(T) = \frac{\phi_0}{2\cdot\pi\cdot\xi^2(0)} \left(\frac{1.77 - 0.43\left(\frac{T}{T_c}\right)^2 + 0.07\left(\frac{T}{T_c}\right)^4}{1.77}\right)^2 \times \left[1 - \frac{1}{2k_BT}\int_0^\infty \frac{d\varepsilon}{\cosh^2\left(\frac{\sqrt{\varepsilon^2 + \Delta^2(T)}}{2k_BT}\right)}\right] \quad (5)$$

where the amplitude of temperature dependent superconducting gap, $\Delta(T)$, is given by [28,29]:

$$\Delta(T) = \Delta(0) \times \tanh\left[\frac{\pi k_B T_c}{\Delta(0)} \sqrt{\eta \frac{\Delta C}{C}\left(\frac{T_c}{T} - 1\right)}\right] \quad (6)$$

where $\eta = 2/3$ for *s*-wave superconductors.

The fit converged with a high quality (with goodness of fit $R = 0.9976$) (Fig. 2). Deduced parameters are: $\xi(0) = 2.33 \pm 0.02\ nm$, $\Delta(0) = 20.2 \pm 1.3\ meV$, $\frac{2\Delta(0)}{k_BT_c} = 4.0 \pm 0.2$, $\frac{\Delta C}{C} = 1.68 \pm 0.15$. Considering that the weak coupling limits of the BCS theory [28-30] are: $\frac{2\cdot\Delta(0)}{k_B\cdot T_c} = 3.53$ and $\frac{\Delta C}{C} = 1.43$, and the upper limits for low-$T_c$ electron-phonon mediated superconductors are: $\frac{2\cdot\Delta(0)}{k_B\cdot T_c} = 5.2$ (for Pb$_{0.50}$Bi$_{0.50}$ alloy [31]) and $\frac{\Delta C}{C} = 3.0$ (for Pb$_{0.70}$Bi$_{0.30}$ alloy [31]), one can conclude that La$_{1-x}$Nd$_x$H$_{10-y}$ (x = 0.09; $P = 180\ GPa$) is moderately strong coupled superconductor.

Final characterization of the superconducting properties of the La$_{1-x}$Nd$_x$H$_{10-y}$ (x = 0.09; $P = 180\ GPa$) was to position this hydride in the empirical Uemura plot [32,33], where heavy fermions, fullerenes, cuprates, pnictides, and hydrogen-rich superconductors [34] form a narrow band exhibited the ratio of the superconducting transition temperature, $T_c$, to the Fermi temperature, $T_F$, within a range:

$$0.01 \lesssim \frac{T_c}{T_F} \lesssim 0.05, \tag{7}$$

while all low-$T_c$ conventional superconductors have much smaller $\frac{T_c}{T_F}$ ratio:

$$\frac{T_c}{T_F} \lesssim 0.001 \tag{8}$$

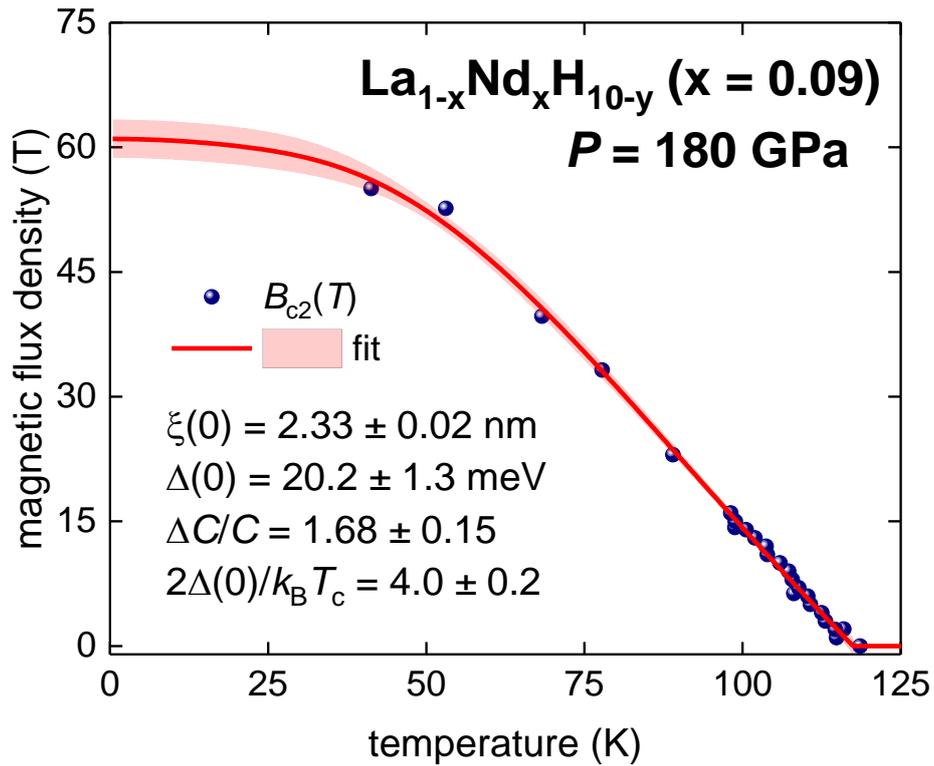

**Figure 2.** Superconducting upper critical field data, $B_{c2}(T)$, and data fit to Eq. 5 for highly compressed $La_{1-x}Nd_xH_{10-y}$ (x = 0.09; P = 180 GPa). Raw $R(T,B)$ dataset is freely available online by Semenok *et al* [14,27]. Deduced parameters are: $\xi(0) = 2.33 \pm 0.02\ nm$, $T_c = 117.5 \pm 0.6\ K$, $\Delta(0) = 20.2 \pm 1.3\ meV$, $\Delta C/C = 1.68 \pm 0.15$, $\frac{2\Delta(0)}{k_B T_c} = 4.0 \pm 0.2$. Fit quality is 0.9976. 95% confidence bands are shown by pink shadow areas.

The Fermi temperature can be calculated by following equation [35]:

$$T_F = \frac{\pi^2}{8 \cdot k_B} \times \left(1 + \lambda_{e-ph}\right) \times \xi^2(0) \times \left(\frac{2\Delta(0)}{\hbar}\right)^2, \tag{9}$$

where all parameters we deduced above. In the result, calculated Fermi temperature is $T_F = 4430 \pm 50\ K$ and, thus, $\frac{T_c}{T_F} = 0.027$ for this superhydride. In the result, $La_{1-x}Nd_xH_{10}$ (x = 0.09;

$P = 180\ GPa$) falls into unconventional superconductors band in the Uemura plot (Fig. 3), and it is located in close proximity to YBa$_2$Cu$_3$O$_{7-\delta}$ and Bi$_2$Sr$_2$Ca$_2$Cu$_3$O$_{11}$ cuprates and other NRTS counterparts.

To address a possible question that the electron-phonon mediated materials are located at the unconventional superconductors band, we should point out that superhydrides are not the only known electron-phonon mediated superconductors which are located in the unconventional superconductors band in the Uemura plot (for instance, we can mention bulk *s*-wave A$_3$C$_{60}$ (A = K, Rb) superconductors).

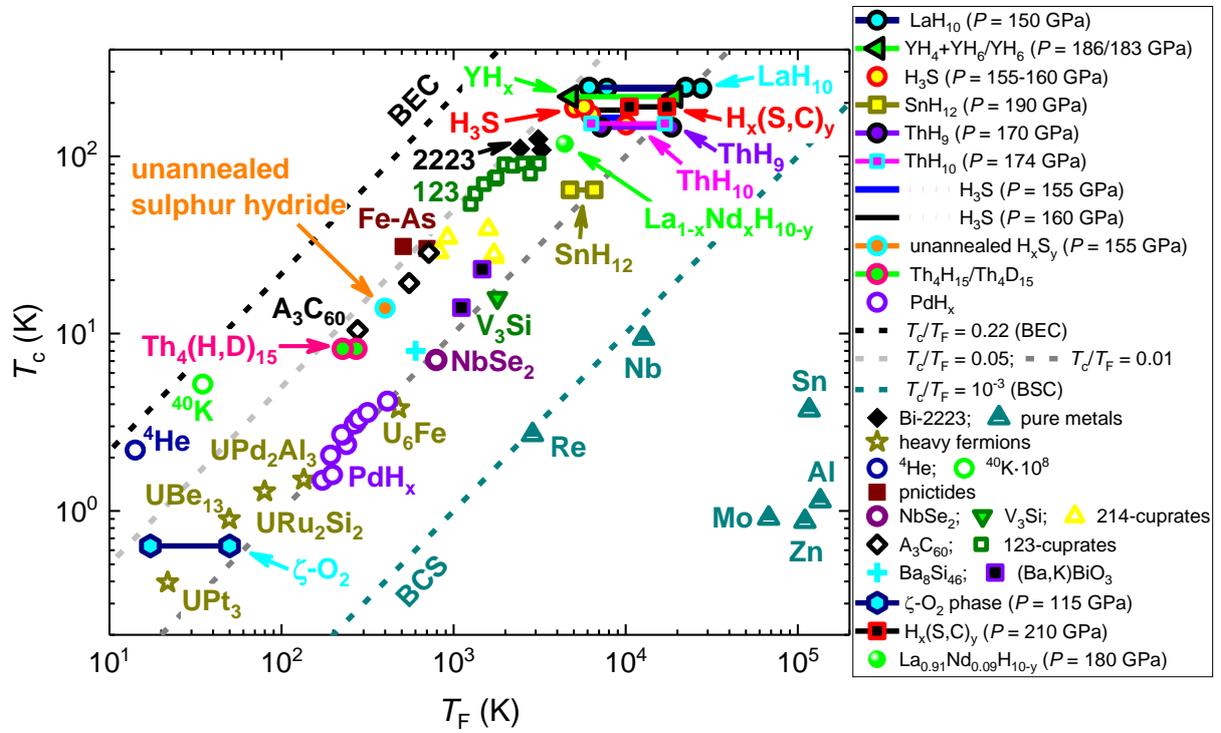

**Figure 3.** Uemura plot ($T_c$ vs $T_F$), where the La$_{1-x}$Nd$_x$H$_{10-y}$ (x = 0.09; $P = 180\ GPa$) compound is shown together with other superconducting families: metals, heavy-fermions, pnictides, cuprates, and near-room-temperature superconductors. Reference on original data can be found in Ref. 34.

From other hand, the Uemura plot is not intended to reveal the pairing mechanism, but rather indicate the geometrical ratio of the characteristic size of the Cooper pair (which is proportional to $\xi(0)$) with average spatial distance between centers of the Cooper pairs. This implies that materials with low superfluid density trend to locate closer to the Bose-Einstein condensate (BEC) line, $\frac{T_c}{T_F} = 0.22$, while materials with high volume concentration of Cooper pairs trend to be located closer to the pure metals, like aluminium for which $\frac{T_c}{T_F} \sim 10^{-5}$. This limit also known as BCS limit. Detailed studies of the transition of one material from BEC into BCS under high pressure can be found elsewhere [36].

To summarise our findings in this work, we can mention that, while the detection of the superconductivity in elemental highly compressed hydrogen is ongoing task [37-39], the near-room temperature superconductivity has observed in several superhydires [1-7]. Here, we analysed experimental magnetoresistance data, $R(T,B)$, for highly compressed La$_{1-x}$Nd$_x$H$_{10-y}$ (x = 0.09; $P = 180\ GPa$) superconductor in which the superconducting order parameter was supressed by magnetic rare earth element (neodymium) impurity. Raw experimental $R(T,B)$ datasets for La$_{1-x}$Nd$_x$H$_{10-y}$ (x = 0.09; $P = 180\ GPa$) was recently reported by Semenok *et al* [14,27]. Deduced parameters, for instance, the gap-to-transition temperature ratio, $\frac{2\Delta(0)}{k_B T_c} = 4.0 \pm 0.2$, and the relative jump in specific heat at transition temperature, $\frac{\Delta C}{C} = 1.7 \pm 0.1$, indicate that La$_{1-x}$Nd$_x$H$_{10-y}$ (x = 0.09; $P = 180\ GPa$) is moderately strong coupled superconductor. This hydride exhibits the ratio of the superconducting transition temperature to the Fermi temperature of $\frac{T_c}{T_F} = 0.027$, and it falls into unconventional superconductors band in the Uemura plot.

**Acknowledgement**

The author thanks Dmitrii V. Semenok (Skolkovo Institute of Science and Technology) and co-workers of Refs. 14,27 for making raw experimental data is freely available prior the peer-review publication of their paper.

**Data availability statement**

The data that support the findings of this study are available from the corresponding author upon reasonable request.

**Declaration of interests**

The author declares that he has no known competing financial interests or personal relationships that could have appeared to influence the work reported in this paper.